\documentclass[aps,prl,reprint,superscriptaddress,showpacs]{revtex4-1} 
\usepackage{graphicx}  
\usepackage{dcolumn}   
\usepackage{amssymb}
\usepackage{physics}
\usepackage{natbib}
\usepackage{color}
\usepackage{hyperref}

\usepackage[textsize=tiny]{todonotes}

\begin{document}

\title{Controlled DC Monitoring of a Superconducting Qubit}

\author{A.~Kringh{\o}j}
\affiliation{Microsoft Quantum Lab Copenhagen and Center for Quantum Devices, Niels Bohr Institute, University of Copenhagen,
Universitetsparken 5, 2100 Copenhagen, Denmark}
\author{T.~W.~Larsen}
\affiliation{Microsoft Quantum Lab Copenhagen and Center for Quantum Devices, Niels Bohr Institute, University of Copenhagen,
Universitetsparken 5, 2100 Copenhagen, Denmark}
\author{B.~van~Heck}
\affiliation{Microsoft Quantum, Station Q, University of California, Santa Barbara, California 93106-6105, USA}
\affiliation{Microsoft Quantum Lab Delft, Delft University of Technology, 2600 GA Delft, The Netherlands}
\author{D.~Sabonis}
\author{O.~Erlandsson}
\author{I.~Petkovic}
\affiliation{Microsoft Quantum Lab Copenhagen and Center for Quantum Devices, Niels Bohr Institute, University of Copenhagen,
Universitetsparken 5, 2100 Copenhagen, Denmark}
\author{D.~I.~Pikulin}
\affiliation{Microsoft Quantum, Station Q, University of California, Santa Barbara, California 93106-6105, USA}
\author{P.~Krogstrup}
\affiliation{Microsoft Quantum Lab Copenhagen and Center for Quantum Devices, Niels Bohr Institute, University of Copenhagen,
Universitetsparken 5, 2100 Copenhagen, Denmark}
\affiliation{Microsoft Quantum Materials Lab Copenhagen, Kanalvej 7, 2800 Lyngby, Denmark}
\author{K.~D.~Petersson}
\author{C.~M.~Marcus}
\affiliation{Microsoft Quantum Lab Copenhagen and Center for Quantum Devices, Niels Bohr Institute, University of Copenhagen,
Universitetsparken 5, 2100 Copenhagen, Denmark}

\begin{abstract}
Creating a transmon qubit using semiconductor-superconductor hybrid materials not only provides electrostatic control of the qubit frequency, it also allows parts of the circuit to be electrically connected and disconnected {\it in situ} by operating a semiconductor region of the device as a field-effect transistor (FET). Here, we exploit this feature to compare {\it in the same device} characteristics of the qubit, such as frequency and relaxation time, with related transport properties such as critical supercurrent and normal-state resistance.  Gradually opening the FET to the monitoring circuit allows the influence of weak-to-strong DC monitoring of a ``live'' qubit to be measured. A model of this influence yields excellent agreement with experiment, demonstrating a relaxation rate mediated by a gate-controlled environmental coupling.  
\end{abstract}

\maketitle

Josephson junctions (JJs) serve as key elements in a wide range of quantum systems of interest for fundamental explorations and technological applications. JJs, which provide the nonlinearity essential for superconducting qubits~\cite{Devoret_2013}, are typically fabricated using insulating tunnel junctions between superconducting metals~\cite{paik_2011}. Alternative realizations using atomic contacts~\cite{bretheau_2013} or superconductor-semiconductor-superconductor (S-Sm-S) junctions~\cite{doh_2005, larsen_2015, delange_2015} are receiving growing attention. Hybrid S-Sm-S JJs host a rich spectrum of new phenomena, including a modified current-phase relation (CPR)~\cite{golubov_2004, Spanton_2017} different from the sinusoidal CPR of metal-insulator-metal tunnel junctions. Other electrostatically tunable parameters include the sub-gap density of states (DOS), shunt resistance~\cite{tinkham}, spin-orbit coupling~\cite{Tosi_2019}, and critical current~\cite{Zou_2017}. 

\begin{figure}[!b]\vspace{-4mm} \includegraphics[width=0.95\columnwidth]{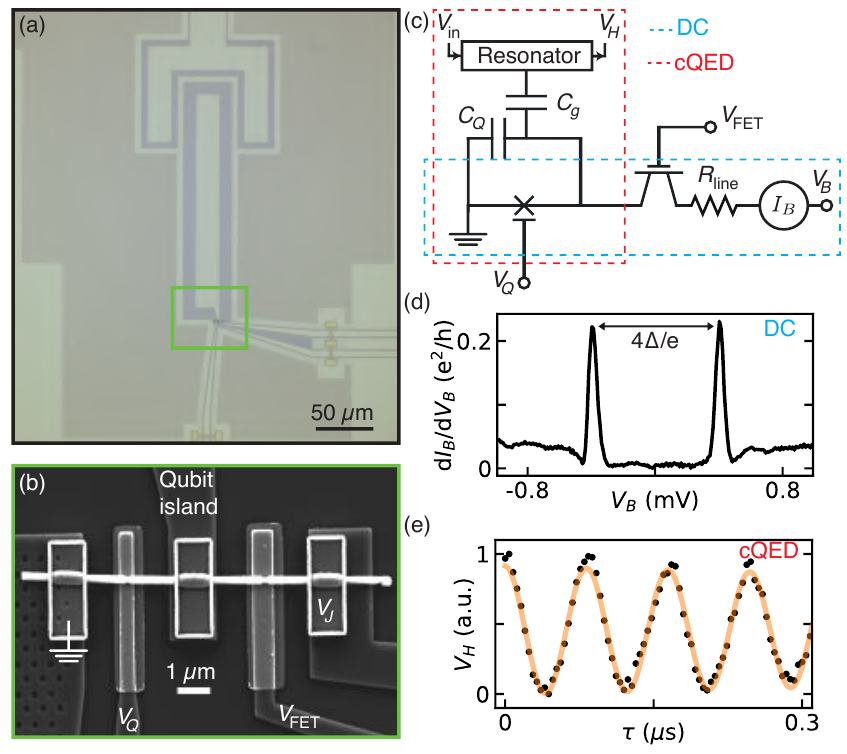}
    \caption{\vspace{-0mm} (a) Optical micrograph of the modified gatemon qubit-device showing the bottom of the readout resonator capacitively coupled to the qubit island. The island is contacted to a nanowire placed in the highlighted green square. (b) Scanning electron micrograph (SEM) of the nanowire in the green rectangle in (a). Two removed segments of the Al shell form the qubit JJ (125\,nm) and the FET (175\,nm), controlled by gates $V_Q$ and $V_\text{FET}$. The bias voltage across the nanowire is indicated $V_J$. (c) Device circuit with FET off for cQED (dashed red box), and FET on allowing transport (dashed blue box). The bias voltage $V_B$ refers to the total voltage drop across both the nanowire and line resistance $R_\text{line}$. (d) Differential conductance d$I_B/$d$V_B$ as a function of bias voltage $V_B$ shows the superconducting gap $\Delta$ of the qubit JJ, with $V_\text{FET}=+4\,$V and $V_Q = -2.9\,$V. (e) Rabi oscillations of the qubit seen in resonator output $V_H$ as a function of drive time $\tau$ at $V_\text{FET}=-3\,$V and $V_Q = -2.5\,$V, with exponentially damped sinusoid (orange).
        }
    \label{device}
\end{figure}

Recent work on S-Sm-S JJs in various platforms relies on either DC (direct current) transport~\cite{Spanton_2017, Goffman_2017} or cQED (circuit quantum electrodynamics) measurements~\cite{Anharmonicity, Hays_2018, Casparis_2018, Wang_2019}. Common to these experiments is that valuable device information is only accessible in one of the two measurement techniques. For instance, measurements estimating individual transmission eigenvalues~\cite{riquelme_2005} or measurements probing the local DOS are directly accessible with DC transport but not with cQED. The prospect of combining these techniques potentially allows a deeper understanding of JJ-based quantum systems.


In this Letter, we investigate a modified S-Sm-S JJ design of a gatemon qubit that combines DC transport and coherent cQED qubit measurements. The device is realized in an InAs nanowire with a fully surrounding epitaxial Al shell by removing the Al layer in a second region (besides the JJ itself) allowing that region to function as a field-effect transistor (FET). By switching the FET between being conducting (``on'') or depleted (``off'') using a gate voltage, we are able to implement a controlled transition between the transport and cQED measurement configurations. We demonstrate that the additional tunability does not compromise the quality of the qubit in the cQED configuration, where the FET is off. 
We further demonstrate control of the qubit relaxation as the FET is turned on, continuously increasing the coupling of the junction to the environment, in agreement with a simple circuit model.
Finally, we demonstrate strong correlation between cQED and transport data by comparing the measured qubit frequency spectrum with the switching current directly measured {\it in situ}. 

Devices were fabricated on a high resistivity silicon substrate covered with a 20\,nm NbTiN film. The nanowire region, qubit-capacitor island, electrostatic gates, on-chip gate-filters, readout resonator, and transmission line were patterned by electron-beam lithography and defined by reactive-ion etching techniques, see Fig.~\ref{device}(a).  The full-shell InAs/Al epitaxial hybrid nanowire is placed at the bottom of the qubit island, see Fig.~\ref{device}(b)~\cite{krogstrup_2015}. Two gateable regions are formed by selective wet-etching of the Al in two $\sim150\,$nm segments defined by electron-beam lithography, aligned with two independent bottom gates, which are separated from the nanowire by a 15\,nm thick HfO$_2$ dielectric. The three superconducting segments---ground, qubit island with capacitance $C_Q$, and DC bias $V_J$---are then contacted with $\sim200$\,nm sputtered NbTiN, see Fig.~\ref{device}(b).  In this circuit, when the FET is on, DC current/voltage measurements are available [blue box in Fig.~\ref{device}(c)]. Depleting the FET allows the device to operate as a qubit, where measurements of the heterodyne demodulated transmission $V_H$ allow qubit state determination and $V_Q$ allows tuning the qubit frequency $f_{01}$ over several GHz [red box in Fig.~\ref{device}(c)].

Setting the voltage on the FET gate to $V_{\text{FET}}=+4\,$V, which turned the FET fully conducting, and the voltage on the qubit JJ to $V_Q = -2.9\,$V makes the voltage drop predominantly across the qubit JJ. In this configuration, the differential conductance d$I_B/$d$V_B$, probes the convolution of the DOS on each side of the JJ, see Fig.~\ref{device}(d). Keeping in mind a simple model of JJ spectroscopy~\cite{tinkham}, we interpret the distance between the two peaks in d$I_B/$d$V_B$ as 4$\Delta$/e = $4\times190\,\mu$V, where $\Delta$ is the induced superconducting gap. In the cQED configuration, with $V_{\text{FET}}=-3\,$V and $V_Q=-2.5\,$V, coherent Rabi oscillations are observed by varying the duration $\tau$ of the qubit drive tone at the qubit frequency $f_{01}=4.6\,$GHz. Following the drive tone, a second tone was applied at the readout resonator frequency, $f_R\sim5.3\,$GHz, to perform dispersive readout where $V_H$ is measured, see Fig.~\ref{device}(e). These experiments are carried out in a dilution refrigerator with a base temperature of $\sim10\,$mK using standard lock-in and DC techniques for the transport measurements and using heterodyne readout and demodulation techniques for the cQED measurements~\footnote{See Supplementary Material for additional details on the experimental setup.}. 


\begin{figure}
    \centering
        \hspace{-2mm}\includegraphics[width=0.95\columnwidth]{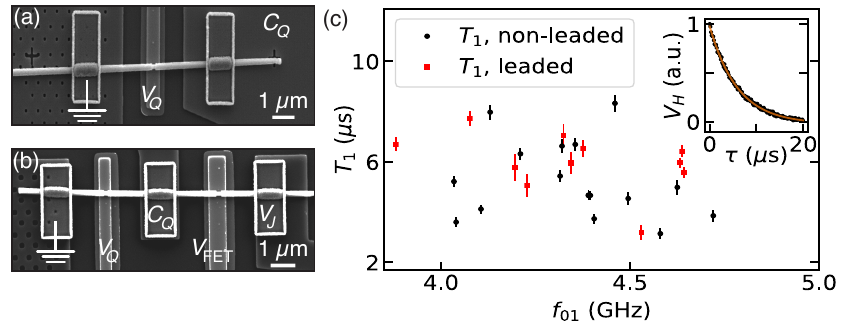}\vspace{-4mm}
    \caption{(a) Scanning electron micrograph of a gatemon without transport lead. $C_Q$ is the capacitance of the qubit island. (b) Same as (a) for gatemon with transport lead, with voltage bias $V_{J}$. (c) Qubit relaxation times $T_1$ of the gatemons as a function of qubit frequency, $f_{01}$. No overall difference between leaded (black circle) and non-leaded (red square) devices are observed. (Inset) Relaxation time $T_{1}$ (black points) at $f_{01}=4.6\,$GHz for the non-leaded device as a function of wait time, $\tau$, with exponential fit (orange curve) yielding $T_1=6\:\mu$s. Error bars are estimated from fit uncertainties. 
    }
    \label{LeadvsNonlead}\vspace{-4mm}
\end{figure}
Having demonstrated the ability to probe the qubit JJ with both transport and cQED techniques, we next compare performance to a nominally identical gatemon without the FET and extra DC lead. Scanning electron micrographs of the two devices are shown in Figs.~\ref{LeadvsNonlead}(a,b). The measured relaxation time, $T_1$, is shown for a range of qubit frequencies, $f_{01}$, controlled by $V_Q$, in Fig.~\ref{LeadvsNonlead}(c). The $T_1$-measurements were carried out by applying a $\pi$-pulse, calibrated by a Rabi experiment at $f_{01}$, followed by a variable wait time $\tau$ before readout, see Fig.~\ref{LeadvsNonlead}(c) inset. Values for $T_1(V_Q)$ were then extracted by fitting $V_H(\tau)$ to an exponential. We observe no systematic difference between the devices, demonstrating that the addition of a transport lead does not compromise the performance in the cQED configuration. 

%
\begin{figure}
    \centering
        \hspace{-2mm}\includegraphics[width=0.95\columnwidth]{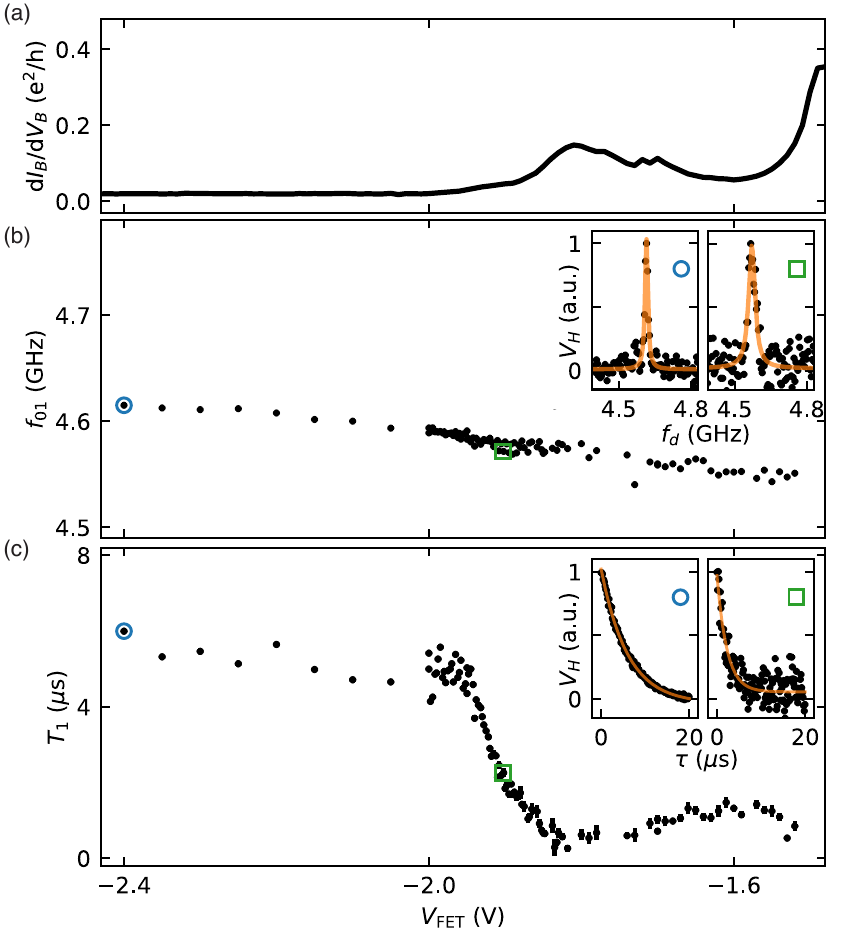}\vspace{-4mm}
    \caption{(a) Differential conductance, d$I_B/$d$V_B$, as a function of FET gate voltage, $V_\text{FET}$, at high bias $V_B = 1.0\,$mV, to approximate normal-state resistance. (b) Qubit frequency $f_{01}$ as a function of $V_\text{FET}$ using two-tone spectroscopy. (Insets) Lorentzian fits (orange) to data points in the main panel as indicated by the corresponding markers (blue circle, green square). From each $V_H$-measurement we subtract an average and normalize to the maximal value. (c) Similar to (b) relaxation times $T_{1}$ from exponential fits (Insets). Error bars are estimated from fit errors.
    }
    \label{FET}\vspace{-4mm}
\end{figure}

We next monitored d$I_B/$d$V_B$, $f_{01}$ and $T_1$ as $V_\text{FET}$ was varied from off (cQED regime) to on (transport regime). The d$I_B/$d$V_B$-measurement shown in Fig.~\ref{FET}(a) illustrates how the FET was turned conducting as $V_\text{FET}$ was increased.  Qubit frequency $f_{01}$ was measured by two-tone spectroscopy, where a drive tone with varying frequency $f_d$ was applied for $2\,\mu$s, followed by a readout tone at $f_R$. A Lorentzian fit is used for each value of $V_\text{FET}$ to extract $f_{01}$, see Fig.~\ref{FET}(b) insets. We attribute the weak dependence of $f_{01}$ on $V_\text{FET}$ to crosstalk between the two gates.

Following each spectroscopy measurement, a $T_1$ measurement was immediately carried out, see Fig.~\ref{FET}(c), yielding a nearly gate-independent value $T_1\sim6\,\mu$s for $V_\text{FET}<-2\,$V. At $V_\text{FET}\sim-2\,$V, we observe a sudden drop in $T_1$, followed by a short revival at $V_\text{FET}\sim-1.8\,$V. We associate the revival in $T_1$ with the corresponding drop in d$I_B/$d$V_B$ observed in Fig.~\ref{FET}(a). For $V_\text{FET}>-1.5\,$V, $f_{01}$ and $T_1$ can no longer be resolved, consistent with the increasing d$I_B/$d$V_B$-values. We note that the d$I_B/$d$V_B$-curve in Fig.~\ref{FET}(a) was shifted horizontally by a small amount (0.1\,V) to align features in d$I_B/$d$V_B$ with corresponding features in $T_1$. This was done to account for hysteresis in the gate sweep, as the cQED and transport measurements were performed sequentially with a large voltage swing on $V_\text{FET}$ of $\sim3\,$V between the two measurements. 


\begin{figure}[b]
    \centering
        \hspace{-2mm}\includegraphics[width=0.95\columnwidth]{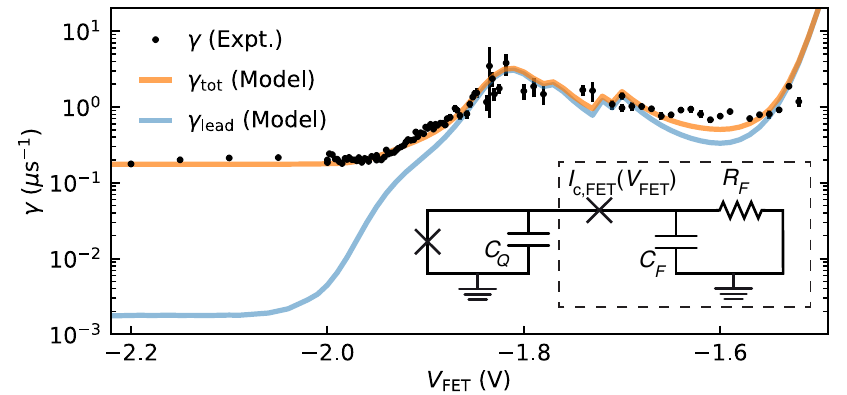}\vspace{-4mm}
    \caption{Relaxation rate $\gamma = 1/T_1$ (black circles) as a function of FET voltage, $V_{\rm FET}$, by inverting the experimental data from Fig.~\ref{FET}(c). Model relaxation rates $\gamma_\text{lead}$ due only to the transport lead (blue) and $\gamma_\text{tot}$ (orange) including lead and non-lead contributions (see text). The circuit model is sketched in the inset where the qubit is coupled to the environment by an effective impedance $Z_\text{env}=i\omega L_\text{FET}+\left(1/R_F+i\omega C_F\right)^{-1}$. The dashed rectangle indicates the environment circuit. 
    }
    \label{model}\vspace{-4mm}
\end{figure}

We develop a circuit model of qubit relaxation in the leaded device. Within the model, the qubit circuit is coupled through the FET to a series resistance $R_F$ and a parallel capacitance $C_F$ representing an on-chip filter on the lead~\cite{mi_2017}. The coupling to the environment via the (superconducting) FET junction is modelled as a gate tuneable Josephson inductance $L_\text{FET}$, giving a total environment impedance $Z_\text{env}=i\omega L_\text{FET}+\left(1/R_F+i\omega C_F\right)^{-1}$. This impedance can be viewed as a single dissipative element with resistance given by
\begin{align}
R_\text{env}=1/\text{Re}[Y]=L_\text{FET}^2\left(R_F^2C_F^2\omega^4+\omega^2\right)/R_F\notag\\+R_F\left(1-2L_\text{FET}C_F\omega^2\right),
\label{eq:R_d}
\end{align}
with admittance $Y=1/Z_\text{env}$~\cite{houck_2008}.
The relaxation rate associated with the lead is given by $\gamma_\text{lead}=1/R_\text{env}C_Q,$ yielding a total decay rate
$\gamma_\text{tot}=\gamma_\text{nolead}+\gamma_\text{lead},$
where $\gamma_\text{nolead}$ is the decay rate associated with relaxation unrelated to the lead.

We estimate $L_\text{FET}=\hbar/2eI_{c,\text{FET}}$~\cite{girvin_2014}, where $I_{c,\text{FET}}$ is the critical current of the FET, which we in turn relate to the normal-state resistance $R_{n,\text{FET}}$ via the relation $I_{c,\text{FET}}R_{n,\text{FET}}=\pi\Delta/2e$~\cite{Ambegaokar_1963}, yielding
\begin{align}
L_\text{FET}=\hbar R_{n,\text{FET}}/\pi\Delta. 
\label{eq:L_FET}
\end{align}
$R_{n,\text{FET}}$ can be found from d$I_B/$d$V_B$ in Fig.~\ref{FET}(a) by subtracting the voltage drop across the line resistance, $R_\text{line}=57\,$k$\Omega$, found by fully opening both the qubit JJ and the FET. To associate this value with $R_{n,\text{FET}}$, we assume no voltage drop across the qubit JJ, justified by $I_{c,\text{FET}}<I_c$, where $I_c$ is the critical current of the qubit JJ.
From electrostatic simulations we estimate $C_Q=38\,$fF~\cite{comsol}. We take $\omega=2\pi\overline{f_{01}}$, where $\overline{f_{01}}=4.6\,$GHz is the average $f_{01}$ in Fig.~\ref{FET}(b), and $\Delta=190\,\mu$eV from Fig.~\ref{device}(d).
Combining Eqs.~\ref{eq:R_d} and \ref{eq:L_FET} with the measured $1/T_1$-data, yields the $\gamma_\text{lead}$-result in Fig.~\ref{model} using $R_F=R_\text{line}$ and $C_F=0.1\,$pF as the best fit parameter. We note that electrostatic simulations give $C_F\sim0.5\,$pF, in reasonable agreement with the best fit value. We define $\gamma_\text{nolead}=1/T_1^\text{mean}$, where $T_1^\text{mean}=5.8\,\mu$s is the mean value of the $T_1$  at $V_\text{FET}<-2\,$V. Using this estimate for $\gamma_\text{nolead}$, we calculate the total relaxation time based on the transport data (orange line in Fig.~\ref{model}) showing excellent agreement with the measured values.
The $T_1$-limit based on the contribution of the lead saturates at $T_{1,\text{lead}}\sim1\,$ms, indicating that leaded gatemon devices can accommodate large improvements in gatemon relaxation times.

\begin{figure}
    \centering
        \includegraphics[width=0.95\columnwidth]{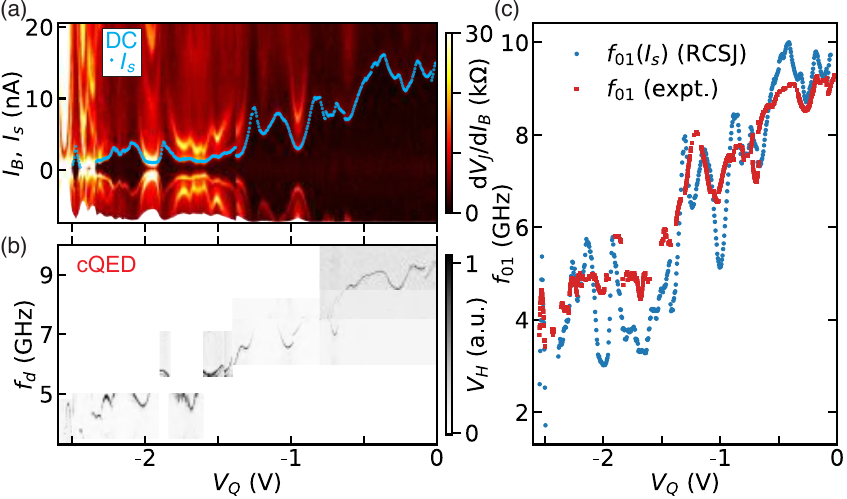}\vspace{-2mm}
    \caption{\vspace{-0mm} (a)  Differential resistance of the qubit JJ, d$V_J/$d$I_B$, as a function of current bias $I_B$ and qubit gate voltage, $V_Q$. Switching current, $I_s$, (blue points) from the edge of the zero-resistance state for increasing sweep at $V_\text{FET}=+4\,$V to turn the FET conducting. (b) Qubit frequency, $f_{01}$, from two-tone spectroscopy as a function of $V_Q$, acquired at $V_\text{FET}=-3\,$V to deplete the FET. The area of missing data at 5.0--5.6\,GHz is due to $f_{01}$ crossing the resonator frequency, $f_R$.  (c) Correlation between transport and cQED data. $f_{01}$ from (b) (red) extracted as in Fig.~\ref{FET}(b) (inset). $f_{01}$ from $I_c$ (blue) extracted by applying an RCSJ model to the data in (a) (see text).
    }
    \label{fig:Isw_f01}
\end{figure}

Combining transport and cQED measurements allows for the correlation between critical current $I_c(V_{Q})$ and $f_{01}(V_{Q})$ to be observed directly. The critical current $I_c$ is extracted from d$I_B/$d$V_B$ and $I_B$ while sweeping $V_B$ and $V_Q$. We extract the voltage drop and differential resistance across the qubit junction, $V_J$ and d$V_J/$d$V_B$, by inverting d$I_B$/d$V_B$ and subtracting $R_\text{line}$. In doing this we assume that there is no voltage drop across the FET junction, since $I_c< I_{c,\text{FET}}$.  The qubit resonance $f_{01}$ is measured over the same $V_Q$ range using two-tone spectroscopy, see Fig.~\ref{fig:Isw_f01}(b). We note that the two-photon transition to the next harmonic is also observed for some $V_Q$-values, visible at a slightly lower frequency than $f_{01}$, given by the anharmonicity.

The relation between the two measurements is shown in Fig.~\ref{fig:Isw_f01}(c). 
In order to estimate $I_c(V_Q)$ we first extract the switching current $I_s(V_Q)$ from the data, taken as the $I_B$-value at which d$V_J/$d$I_B$ is maximal, while sweeping $I_B$ from negative to positive values [blue dots in Fig.~\ref{fig:Isw_f01}(a)]. Bright features at high bias ($I_B>I_s$) are likely associated with multiple Andreev reflection (MAR)~\cite{klapwijk_1982}. To extract $I_c$ from the measured $I_s$, we model the qubit JJ as an underdamped RCSJ junction with a sinusoidal current-phase relation $I=I_c\sin\phi$. Furthermore, we note the small difference between the return current $I_r$ (same definition as $I_s$ at negative $I_B$) is slightly smaller than $I_s$~\cite{supplement_critical}. In this case, $I_s$ corresponds to the current of equal stability between the resistive and non-resistive state~\cite{kautz_1990}. Under this condition, and for large quality factors, $Q\gg1$, the ratio $I_{s}/I_{c}$ depends on quality factor $Q=R\sqrt{2eI_cC_Q/\hbar }$ as
\begin{align}
I_{s}/I_{c}=(2+4/\pi)Q^{-1}+(2+\pi)Q^{-2},
\label{eq:km_eq}
\end{align}
where $R=\left(1/R_J+1/R_\text{line}\right)^{-1}$ and $R_J$ is the shunt resistance~\cite{kautz_1990}. 
For simplicity, we take $R_J$ to be equal to the normal state resistance of the junction, $R_N$.
We then apply the Ambegaokar-Baratoff relation $I_cR_J=\pi\Delta/2e$~\cite{Ambegaokar_1963}, which allows us to extract $I_c$ by inverting Eq.~\ref{eq:km_eq} numerically~\footnote{Numerical code and data accompanying the analysis of Fig.~\ref{fig:Isw_f01}(c) is found at: \url{https://github.com/anderskringhoej/dc_qubit}.}. The extracted values of $I_c$, in turn, yield values for $Q$ in the range 10-20, consistent with our initial assumptions.
For these values of $Q$, the RCSJ model takes the electron temperature to be $>50$\,mK to account for the weak asymmetry in $I_s$ and $I_r$~\cite{supplement_critical}.
Finally, we relate $I_c$ to $f_{01}$ by using the numerical solution of the standard transmon Hamiltonian, $H=4E_C(n-n_g)^2-E_J\cos(\phi)$~\cite{koch_2007}, with $E_J=\hbar I_c/2e$ and $E_C/h=e^2/2hC_Q=512\,$MHz, at the charge degeneracy point with offset charge $n_g=0.5$.

A comparison of the measured and estimated $f_{01}$ is shown in Fig.~\ref{fig:Isw_f01}(c). The model (RCSJ) curve is shifted horizontally by 0.05\,V to align the features at $\sim-2.5\,$V and can be attributed to small gate hysteresis. A clear correlation is observed between the two measurement techniques, especially evident from the matching of local minima and maxima of both spectra and the overall agreement of the absolute values.
We attribute the residual quantitative discrepancy to the simplifying assumptions used to determine the shunt resistance of the RCSJ model, which likely do not capture the possible gate dependence of the subgap DOS of the qubit JJ.
In addition, the assumption of sinusoidal CPR will break down as the qubit JJ is opened due to increasing mode transmission in the semiconductor junction, leading to small overshoots of the model as perhaps seen around $V_Q\sim 0\,$V.


In summary, we have demonstrated the compatibility of DC transport and cQED measurement techniques in gatemon qubits. This method may extend to other material platforms such as two-dimensional electron gases~\cite{Casparis_2018} or graphene~\cite{Wang_2019, Kroll_2018}. Furthermore, we achieve a controllable relaxation rate potentially relevant for a range of qubit applications such as tunable coupling schemes~\cite{Chen_2014,  Casparis_2019}, and controlled qubit relaxation and reset protocols~\cite{jones_2013, Ma_2019}. In addition, we have demonstrated clear correlation between DC transport and cQED measurements motivating future extensions, such as studying CPRs~\cite{Spanton_2017} or probing channel transmissions by studying multiple Andreev reflections~\cite{Goffman_2017} combined with cQED experiments~\cite{Anharmonicity, Hays_2018, Tosi_2019}. Combining well-established transport techniques in quantum dot physics with qubit geometries may also be an interesting research direction~\cite{DeFranceschi_2010}. Potentially this geometry is also a promising platform to coherently probe Majorana zero modes in cQED measurements~\cite{Ginossar_2014}, as transport signatures have been demonstrated, both in half-shell nanowires~\cite{Mourik_2012} and full-shell wires~\cite{lutchyn_2018, Sole_2018}.

\begin{acknowledgments}
This work was supported by Microsoft and the Danish National Research
Foundation. We acknowledge Robert McNeil, Marina Hesselberg, Agnieszka Telecka, Sachin Yadav, Karolis Parfeniukas, Karthik Jambunathan and Shivendra Upadhyay for the device fabrication. We thank Lucas Casparis and Roman Lutchyn for useful discussions, as well as Ruben Grigoryan for input on the electronic setup. BvH would like to thank the Center of Quantum Devices, Niels Bohr Institute for the hospitality during the time in which this study was carried out.
\end{acknowledgments}

%

\newcommand{\beginsupplement}{%
        \setcounter{table}{0}
        \renewcommand{\thetable}{S\arabic{table}}%
        \setcounter{figure}{0}
        \renewcommand{\thefigure}{S\arabic{figure}}%
     }

\onecolumngrid
\section{Supplementary Material}
\maketitle
\subsection{Experimental setup}
\beginsupplement

\begin{figure}
    \centering
        \includegraphics[width=1\textwidth]{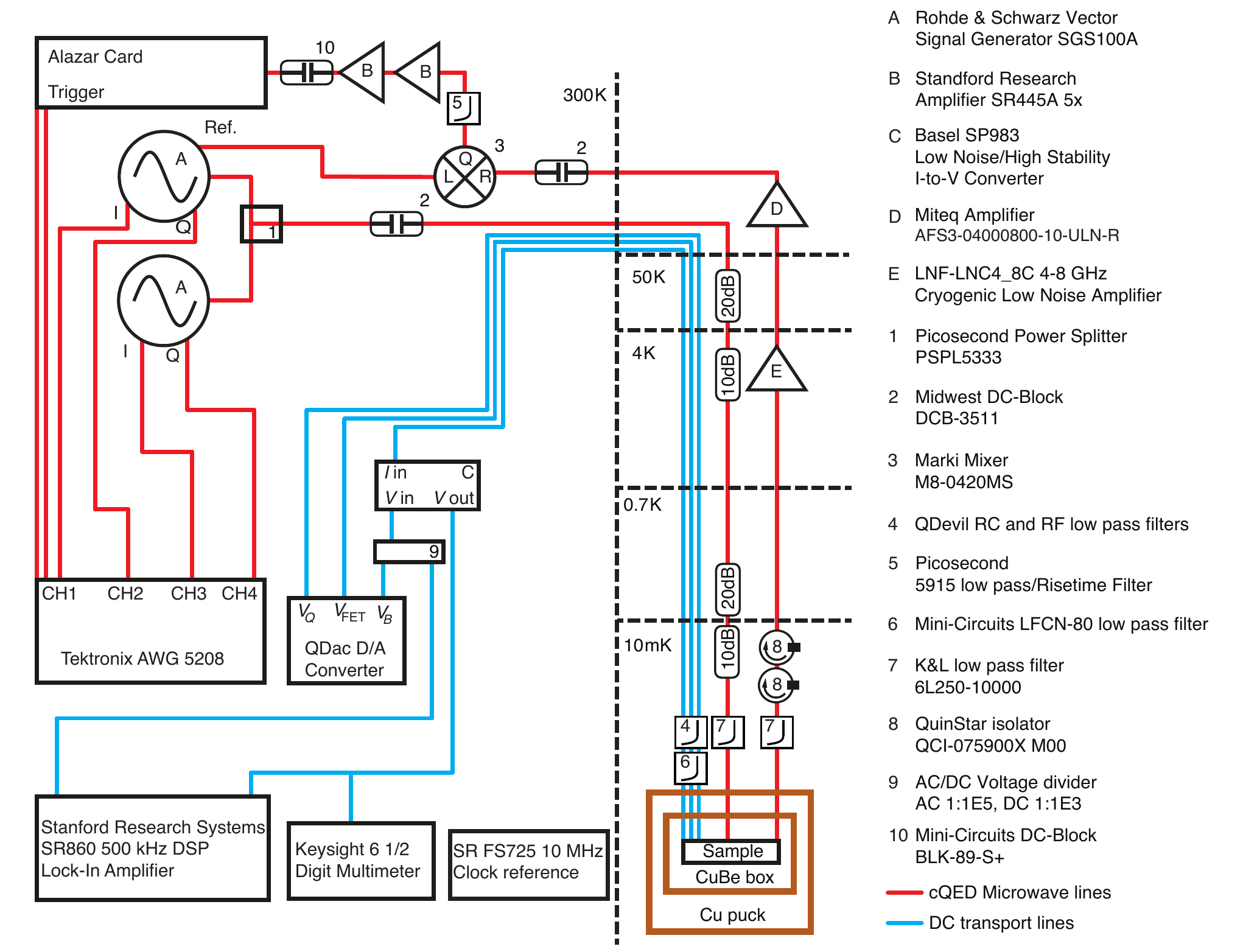}\vspace{-2mm}
    \caption{\vspace{-0mm} Schematic of the setup. Blue lines refer to lines used for the DC transport and red lines are the microwave drive lines used for qubit manipulation and readout. The signal generators, AWG, and Alazar card are all connected to the SR FS725 10\,MHz clock reference for synchronisation.
    }
    \label{setup}
\end{figure}

The measurements presented in the paper are conducted in a cryogen-free dilution refrigerator with a base temperature of $\sim10\,$mK. A detailed schematic of the electronic setup is shown in Fig.~\ref{setup}. The sample is mounted to a Cu circuit board located in a indium sealed CuBe box mounted inside another Cu box, which is thermally attached to the mixing chamber plate. DC lines (blue lines in Fig.~\ref{setup}) connect to the sample through a loom heavily filtered at frequencies above 80\,MHz via both the QDevil and the LFCN-80 low pass filters. For transport measurements we measure a small AC current using the SR860 lock-in amplifier while also measuring the DC current to ground with the Keysight multimeter. Both current signals are amplified and converted to a voltage by the Basel SP983 I-to-V converter.

Two microwave coaxial drive lines connect to the sample (red lines in Fig.~\ref{setup}). The combined input signal is generated by two RF sources and is heavily attenuated and filtered above 10\,GHz with a K$\&$L low pass filter. These two signals are used for qubit drive and readout drive, respectively. The output signal is again filtered and amplified at the 4\,K stage with a cryogenic low noise amplifier with a bandwidth of 4--8\,GHz with further amplification at room temperature using the Miteq amplifier. The output signal is down converted to an intermediate frequency by mixing with a local oscillator and filtering of the high frequency component. After another amplification stage using the SR445A amplifier, the intermediate frequency signal is digitized and digitally down converted in order to extract the in-phase and quadrature components of the readout signal.

The SR FS725 10\,MHz clock reference is connected to the Alazar card, signal generators and the AWG for synchronisation of the microwave signals.

\subsection{RCSJ modelling details and data}
\begin{figure}
    \centering
        \includegraphics[width=0.65\columnwidth]{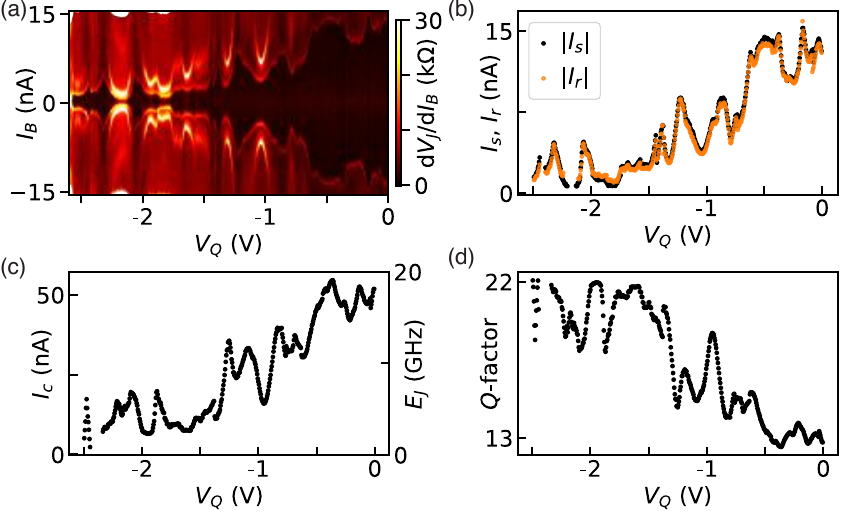}\vspace{-2mm}
    \caption{\vspace{-0mm} (a) DC transport measurement of d$V_J/$d$I_B$ as a function of $I_B$ and $V_Q$, acquired in the same way as the data presented in Fig.~5(a). In this measurement, both the transition to a non-resistive state at negative $I_B$-values and the transition to the resistive state at positive $I_B$-values are observed. (b) Absolute values of the extracted return current $I_r$ and switching current as a function of $V_Q$, illustrating the weak asymmetry in their values. (c) Extracted critical current $I_c$ (left $y$-axis, converted to $E_J=\hbar I_c/2e$ on the right $y$-axis). (d) Extracted quality factor $Q$ from the numerical solutions to Eq.~3. 
    }
    \label{fig:transport}
\end{figure}

To supplement the data and the analysis presented in Fig.~5, we measured d$V_J/$d$I_B$ as a function of $I_B$ and $V_Q$ for a $I_B$-range where we were able to extract both $I_s$ and $I_r$ for the entire $V_Q$-range, see Fig.~\ref{fig:transport}(a). This dataset shows quantitatively almost the same features as the dataset in the main text. However, due to a larger amount of drift, possibly due to longer acquisition time, we use the dataset in Fig.~5(a) to perform the modelling in the main text.
From the measurement shown in Fig.~\ref{fig:transport}(a) we are able to extract both $I_s$ and $I_r$, see Fig.~\ref{fig:transport}(b). Here we observe a weak asymmetry between $I_r$ and $I_s$ for the full $V_Q$-range, which justifies the use of the RCSJ model applied in the analysis of Fig.~5(c).

In addition, we compute the extracted critical current $I_c$ and $E_J=\hbar I_c/2e$ used in our RCSJ analysis, as shown in Fig.~\ref{fig:transport}(c). Based on these $E_J$-values we estimate the electron temperature $T$ to be $>$\,50mK, such that the $k_BT/E_J$-ratios account for the weak asymmetry between $I_r$ and $I_s$~\cite{kautz_1990}. To further justify the application of the $Q\gg1$ limit, we numerically extract the $Q$-values, as shown in Fig.~\ref{fig:transport}(d).

\end{document}